\documentstyle[aas2pp4]{article}

\slugcomment{The Astrophysical Journal, submitted 2000 October 21;
accepted 2001 February 15}

\lefthead{Wanajo et al.}
\righthead{The $r$-PROCESS IN NEUTRINO-DRIVEN WINDS}

\begin{document}

\title{THE $r$-PROCESS IN NEUTRINO-DRIVEN WINDS FROM NASCENT, ``COMPACT''
       NEUTRON STARS OF CORE-COLLAPSE SUPERNOVAE}

\author{\sc Shinya Wanajo$^{1, 2}$,
            Toshitaka Kajino$^{2, 3, 4}$,
            Grant J. Mathews$^{2, 5}$,
            and Kaori Otsuki$^2$}
\affil{\em wanajo@sophia.ac.jp,
       kajino@nao.ac.jp, gmathews@nd.ed, otsuki@th.nao.ac.jp}

\bigskip
\affil{The Astrophysical Journal, submitted 2000 October 21;
accepted 2001 February 15}

\altaffiltext{1}{Department of Physics, Sophia University,
                 7-1 Kioi-cho, Chiyoda-ku, Tokyo, 102-8554, Japan}
\altaffiltext{2}{Division of Theoretical Astrophysics,
                 National Astronomical Observatory,
                 2-21-1 Osawa, Mitaka, Tokyo 181-8588, Japan}
\altaffiltext{3}{Depertment of Astronomy, University of Tokyo,
                 7-3-1 Hongo, Bunkyo, Tokyo 113-0033, Japan}
\altaffiltext{4}{Department of Astronomical Science,
                 The Graduate University for Advanced Studies,
                 2-21-1 Osawa, Mitaka, Tokyo 181-8588, Japan}
\altaffiltext{5}{Center for Astrophysics, Department of Physics,
                 University of Notre Dame, Notre Dame, Indiana 46556, USA}

\begin{abstract}
We present calculations of $r$-process nucleosynthesis in
neutrino-driven winds from the nascent neutron stars of core-collapse
supernovae.  A full dynamical reaction network for both the
$\alpha$-rich freezeout and the subsequent $r$-process is employed. The
physical properties of the neutrino-heated ejecta are deduced from a
general relativistic model in which spherical symmetry and steady flow
are assumed. Our results suggest that proto-neutron stars with a large
compaction ratio provide the most robust physical conditions for the
$r$-process. The third peak of the $r$-process is well reproduced in the
winds from these ``compact'' proto-neutron stars even for a moderate
entropy, $\sim 100-200 N_A k$, and a neutrino luminosity as high as
$\sim 10^{52}$~ergs~s$^{-1}$. This is due to the short dynamical
timescale of material in the wind.  As a result, the overproduction of
nuclei with $A \lesssim 120$ is diminished (although some overproduction
of nuclei with $A \approx 90$ is still evident). The abundances of the
$r$-process elements per event is significantly higher than in previous
studies. The total-integrated nucleosynthesis yields are in good
agreement with the solar $r$-process abundance pattern. Our results have
confirmed that the neutrino-driven wind scenario is still a promising
site in which to form the solar $r$-process abundances.  However, our
best results seem to imply both a rather soft neutron-star equation of
state and a massive proto-neutron star which is difficult to achieve
with standard core-collapse models.  We propose that the most favorable
conditions perhaps require that a massive supernova progenitor forms a
massive proto-neutron star by accretion after a failed initial neutrino
burst.

\end{abstract}


\section{INTRODUCTION}

The $r$-process accounts for the origin of about a half of the
abundances of elements heavier than iron.  The other half mostly come
from the $s$-process. Nevertheless, the astrophysical site for the
$r$-process has not yet been unambiguously identified. Among the large
number of proposed candidates, the neutrino-heated ejecta from a nascent
neutron star (hereafter ``neutrino-driven wind'') has been suggested
(\cite{Woos92}; \cite{Meye92}) as perhaps the most promising
site. Woosley et al. (1994) demonstrated that the solar $r$-process
abundances were well reproduced as material was ablated from the
proto-neutron star in neutrino-driven winds. There are, however, a few
serious problems in their numerical results. First, elements with $A
\sim 90$ were significantly overproduced by over a factor of
100. Second, the requisite high entropy ($\gtrsim 400 N_{\rm A} k$) in
their supernova simulations has not been duplicated by other independent
theoretical studies (Witti, Janka, \& Takahashi 1994; Takahashi, Witti,
\& Janka 1994; Qian \& Woosley 1996). On the other hand, other viable
sites have been demonstrated to also naturally reproduce the solar
$r$-process abundance pattern.  For example, it has been shown that
neutron-star mergers can naturally reproduce the abundances of nuclei
with $A \gtrsim 130$ (Freiburghaus, Rosswog, \& Thielemann
1999). Collapsing O-Ne-Mg cores resulting from progenitor stars of $\sim
8-10 M_\odot$ has also been suggested (Wheeler, Cowan, \& Hillebrandt
1998) to be a promising site for the $r$-process.

In contrast to the above difficulties in the theoretical studies of the
neutrino-driven wind scenario, observational data seem to confirm that
Type~II supernovae are indeed the site for the production of $r$-process
elements.  For example, spectroscopic studies of metal-poor halo stars
indicate very early enrichment of $r$-process elements consistent with
production in massive stars (Sneden et al. 1996, 1998; Ryan, Norris, \&
Beers 1996).  Using a chemical evolution model Ishimaru \& Wanajo (1999)
have furthermore demonstrated that the large dispersion of the
$r$-process element europium in metal-poor halo stars is also reproduced
if the $r$-process originates from Type~II supernovae (see also
Tsujimoto, Shigeyama, \& Yoshii 2000; Qian 2000).

As one way to fix the schematic wind models, it has been suggested that
general relativistic effects increase the entropy and reduce the
dynamical timescale of neutrino-driven wind models.  Using semi-analytic
studies for spherically symmetric, steady flow of neutrino-heated
ejecta, Qian \& Woosley (1996) showed that the inclusion of a first
post-Newtonian correction to the gravitational force equation increased
the entropy by $\sim 60\%$ and reduced the timescale by a factor of
$\sim 2$. Cardall \& Fuller (1997) further developed this argument by
considering a fully general relativistic treatment. They considered a
wide range of neutron-star compaction ratios (ratio of the gravitational
mass to the neutron star radius in Schwarzschild coordinates). They
showed that a more compact neutron star leads to significantly higher
entropy and a shorter dynamical timescale in the wind. In their study,
however, the sensitivity to the neutron star mass and neutrino
luminosity were not considered.

Otsuki et al. (2000, hereafter Paper~I) studied physical conditions of
neutrino winds in a manner similar to that of Cardall \& Fuller (1997),
for wide ranges of the neutron star masses and the neutrino
luminosities. They suggested that high entropy and short dynamical
timescales were obtained for a massive neutron star and a high neutrino
luminosity. However, the radii of the neutron stars were fixed to be
10~km in their study.  Moreover, these studies did not perform
calculations of realistic $r$-process nucleosynthesis except for one
specific case of a neutron star mass ($= 2 M_\odot$) and a neutrino
luminosity ($L_{\nu_e} = 10^{52}$~ergs~s$^{-1}$) in Paper~I.

The purpose of the present study is, therefore, to examine more
quantitatively the $r$-process in neutrino-driven winds utilizing a
fully implicit reaction network of over 3000 isotopes. Trajectories of
thermodynamic quantities for material in the winds are derived for
various combinations of three parameters: the neutron-star mass; radius;
and neutrino luminosity. The semi-analytic, general relativistic model
developed in Paper~I is adopted to describe material in the
neutrino-driven wind.

\begin{deluxetable}{rrrr}
\tablecaption{Model Parameters \label{tab1}}
\tablewidth{0pt}
\tablehead{
\colhead{} & \colhead{A} & \colhead{B} & \colhead{C}}
\startdata
$M/M_\odot$                   & 1.4     & 1.4     & 2.0     \\
$R$ (km)                      & 10      & 7       & 10      \\
$GM/c^2R$                    & 0.21    & 0.30    & 0.30    \\
\enddata
\end{deluxetable}

The paper is organized as follows: In the next section, the dependences
of the entropy and the dynamical timescale on the neutron-star mass,
compaction ratio, and neutrino luminosity are discussed.  In \S~3, the
$r$-process nucleosynthesis is performed using the neutrino-driven wind
trajectories obtained in \S~2. The yields for various neutrino
luminosities are mass-integrated by assuming an exponential time
evolution for the neutrino luminosity. These are compared to the solar
$r$-process abundances. The implications of this study are discussed in
\S~4.

\section{NEUTRINO-DRIVEN WINDS}

\begin{figure}[t]
\epsscale{0.70}
\plotone{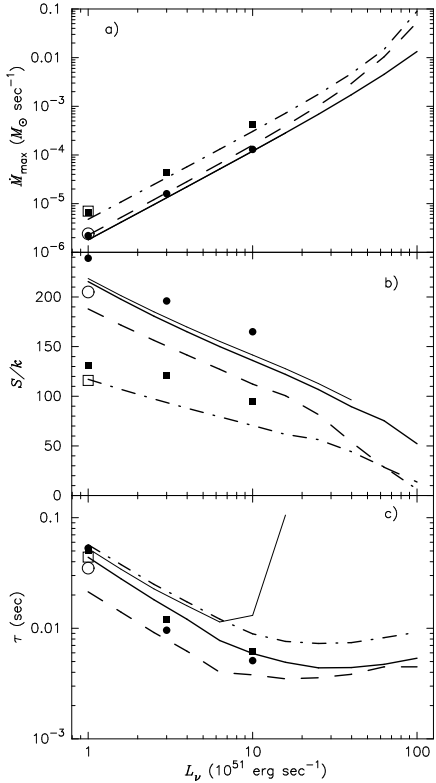}
\caption{\footnotesize (a) The maximum mass ejection rates, (b) entropy
per baryon, and (c) dynamical timescales measured as the time for
material to cool from $T = 0.5$~MeV to 0.2~MeV for models~A (dot-dashed
line), B (dashed line), and C (thick-solid line), as functions of
$L_\nu$ for the supersonic winds. The thin-solid lines in the panels (b)
and (c) are for the subsonic wind with $\dot M = 0.995 \times \dot
M_{\rm max}$ for model~C (see text). Also denoted are the results from
Sumiyoshi et al. (2000, filled squares and circles) and Qian \& Woosley
(1996, open squares and circles). In both cases, the squares and circles
are the results with the same model parameters $M/M_\odot$ and $R$ as
the models A and B, respectively}
\end{figure}

The models of neutrino-driven winds used in this work were developed in
detail in Paper~I.  We briefly describe the models here and point out
some improvements which have been added. The system is treated as time
stationary and spherically symmetric.  The physical variables in the
neutrino wind are then functions of the radius $r$ only. The ejected
mass $M_{\rm ej}$ by neutrino heating is assumed to be negligible
compared to the mass of the neutron star $M$. Therefore, the
gravitational filed in which the neutrino-heated matter moves can be
treated as a fixed background Schwarzschild metric. The equations of
baryon, momentum, and mass-energy conservation then become,
\begin{eqnarray} \label{eqn:wind}
\dot M            & = & 4\pi r^2 \rho u \:, \\ \label{eqn:mdot}
u \frac{du}{dr}   & = & - \frac{1 + (u/c)^2 - 2GM/c^2r}
                        {\rho \left(1 + \varepsilon/c^2 \right) + P/c^2}
                        \frac{dP}{dr}
                        - \frac{GM}{r^2} \:, \\ \label{eqn:dudr}
\dot q            & = & u \left(\frac{d\varepsilon}{dr}
                        - \frac{P}{\rho^2}
                          \frac{d\rho}{dr}\right) \:, \label{eqn:qdot}
\end{eqnarray}
where $\dot M$ is the mass ejection rate from the surface of the neutron
star, $\dot q$ is the heating rate, $\rho$ is the mass density, $P$ is
the pressure, and $\varepsilon$ is the specific internal energy
(\cite{Shap83}). The velocity $u$ is related to the proper velocity of
the matter $v$ as measured by a local, stationary observer by
\begin{equation}
v = \left[1 + (u/c)^2 - 2GM/c^2r \right]^{-1/2} u \:.
\end{equation}

The source term $\dot q$ includes both heating and cooling by neutrino
interactions. Heating is due to neutrino and anti-neutrino captures by
free nucleons, neutrino scattering on electrons and positrons, and
neutrino-antineutrino pair annihilation into electron-positron pairs.
The redshift of neutrino energies and bending of the neutrino
trajectories by general relativity are explicitly taken into
account. Cooling is due to electron and positron capture by free
nucleons, and annihilation of electron-positron pairs into
neutrino-antineutrino pairs.  For the latter, the more accurate table of
Itoh et al. (1996) is utilized rather than equation~(15) in Paper~I.
All other neutrino heating and cooling rates are taken from Paper~I.

The neutrino luminosities $L_\nu$ of all flavors are assumed to be
equal, and the RMS average neutrino energies are taken to be 12, 22, and
34~MeV, for electron, anti-electron, and the other flavors of neutrinos,
respectively. The temperature at the surface of the neutron star is
determined so that heating and cooling by neutrino interactions are in
equilibrium. The surface of the neutron star is arbitrarily defined as
the point at which density has dropped to $\rho =
10^{10}$~g~cm$^{-3}$. The equation of state for the electron and
positron gas includes not only the relativistic pairs as in Paper~I but
also the partially relativistic pairs, which are important during the
$\alpha$-process.  Once $M$, $R$, and $L_\nu$ are specified along with
the boundary condition $\dot M$, we obtain numerically the velocity and
thermodynamic quantities of the neutrino-driven wind as functions of
$r$.

In Paper~I, $\dot M$ for each neutrino-driven wind was fixed such that
the temperature at $r=10,000$~km was 0.1~MeV. However in that case no
physical solution exists for $L_\nu \gtrsim 10^{52}$~ergs~s$^{-1}$,
because the temperature does not cool to 0.1~MeV before the wind reaches
$r=10,000$~km for any physical solutions. Note that equations~(1)-(3)
have no physical solution for $\dot M > \dot M_{\rm max}$. In order to
avoid this discontinuity, we instead adopt $\dot M = \dot M_{\rm max}$,
which allows for a supersonic wind. The advantage is that the wind
solution exists for any $L_\nu$ even larger than
$10^{52}$~ergs~s$^{-1}$.

We evaluate the key parameters for the $r$-process, i.e., entropy and
dynamical timescale, as a function of $\dot M_{\rm max}$ instead of
arbitrary $\dot M$ for each $L_\nu$. When nucleosynthesis calculations
are performed, we actually take $\dot M$ to be slightly smaller than
$\dot M_{\rm max}$ to obtain solutions for subsonic winds, as discussed
below.  As initial conditions, we make the reasonable assumptions that
the electron fraction $Y_ e$ is set to 0.5, and that no
$\alpha$-particles or heavy elements have yet formed.

In this study, we explore three specific models of the neutrino-driven
winds based upon combinations of two parameters $M$ and $R$. The models
given in Table 1 correspond to: A) $(M, R)$ $= (1.4 M_\odot, 10$~ km);
B) $(M, R)$ $= (1.4 M_\odot, 7$~km); and C) $(M, R)$ $= (2.0 M_\odot,
10$~km). The difference between models A and B is the compaction ratio
$GM/c^2R$, while neutron star mass is the same. On the other hand,
effects of general relativity are manifest through the compaction ratio. 
Models B and C therefore involve different masses and radii, while the
compaction ratio is the same. The compaction ratio for models B and C is
rather large $\sim 0.3$. However, this value is still consistent with
relatively soft equations of state for high density matter, or as a
metastable state prior to collapse to a blackhole (\cite{Brow94};
\cite{Baum96}). Note that the case of a smaller compaction ratio
$\lesssim 0.2$ is omitted in this study, where $r$-processing may not be
possible (\cite{Card97}).

Figure~1 shows: a) the mass ejection rate $\dot M_{\rm max}$; b) the
entropy per baryon $S/k$ at $T = 0.5$~MeV; and c) the dynamical
timescale $\tau$ measured as the time for material to cool from $T =
0.5$~MeV to 0.2~MeV, as a function of $L_\nu$ for supersonic winds. The
mass ejection rate is the maximum value for physical solutions of the
winds. The dot-dashed, dashed, and thick-solid lines respectively denote
models A, B, and C. As can be seen in Figure~1, for larger $L_\nu$,
$\dot M_{\rm max}$ and $S$ take larger and smaller values, respectively. 
A similar trend in entropy can be seen in $\tau$. However, the time
scale saturates at $L_\nu \sim$ a few times
$10^{52}$~ergs~s$^{-1}$. This can be traced to the smaller transonic
radii $r_{\rm c}$ and higher initial temperature for higher neutrino
luminosities. The temperature gradient is steeper when $r$ is closer to
the neutron star surface, while being approximately constant for $r >
r_{\rm c}$ (Paper~I). Thus, the temperature decreases more slowly for $r
> r_{\rm c}$ than for $r < r_{\rm c}$. For higher luminosity cases
$\gtrsim 10^{52}$~ergs~s$^{-1}$, the winds pass $r_{\rm c}$ with $T >
0.2$~MeV. As a result, the dynamical timescale does not decrease for
higher luminosities.

The $r$-process favors higher entropy and shorter dynamical
timescale. Thus, a robust $r$-process is difficult to obtain once the
neutrino luminosity is higher than a few times
$10^{52}$~ergs~s$^{-1}$. Comparing the results of models~A and B (or
models~A and C), we find that more ``compact'' proto-neutron stars eject
less material, and obtain significantly higher entropies and shorter
dynamical timescales.  This is in good agreement with the results of
Cardall \& Fuller (1997). Clearly, the more ``compact'' neutron star
models (B and C) will obtain more favorable physical conditions for the
$r$-process.

On the other hand, comparing the results of models~B and C which are at
the same compaction ratio, we see that a more massive (or smaller
radius) proto-neutron star ejects slightly less material and provides
higher entropy, but involves a larger dynamical timescale. These can be
explained by the relations $\dot M \propto R^{5/3}M^{-2}$, $S \propto
R^{-2/3}M$, and $\tau \propto RM$ for fixed energy and neutrino
luminosity (\cite{Qian96}) in the Newtonian limit. Based on these
results alone it is difficult to judge which of the models~B or C is
more favorable for the $r$-process. Differences of the nucleosynthesis
results between these models are to be discussed in the next section.

Figure~1 also shows the results of the numerical simulation of the winds
from Sumiyoshi et al. (2000) for $(M, R)$ $= (1.4 M_\odot, 10$~ km) and
$= (2.0 M_\odot, 10$~ km), respectively. These are denoted by filled
squares and circles.  Open squares and circles show the results of Qian
\& Woosley (1996) calculated with post-Newtonian corrections for the
same parameter sets of $(M, R)$. Our results for $\dot M$, $S$, and
$\tau$ are in good agreement with theirs. Ultimately, mass ejection
rates must be determined by a detailed hydrodynamic simulation of a
proto-neutron star with neutrino transport being taken into
consideration. Nevertheless, our mass ejection rates, entropies, and
dynamical timescales are in good agreement with those of Sumiyoshi et
al. (2000), who calculated steady-flow neutrino-driven winds with the
use of a one-dimensional hydrodynamic code including neutrino heating
and cooling processes. Thus, we conclude that the true mass ejection
rates are probably close to the supersonic winds obtained here.

Without imposing an outer boundary, however, the temperature and density
diminish too quickly in the supersonic winds. This is not favorable for
the $r$-process. Therefore, a slightly smaller mass ejection rate,
$0.995 \times \dot M_{\rm max}$, is adopted for application to
$r$-process calculations in the next section. The advantage of this
reduction is that the temperature and density fall off slowly during the
$r$-process ($T_9 \lesssim 2.5$) in these subsonic winds, ensuring
enough time for the nucleosynthesis of heavy elements to proceed. This
situation resembles the wind with an outer boundary, as if a supersonic
wind interacts with a wall shock to sustain the temperature and density
suitable for the $r$-process. The entropy per baryon and the dynamical
timescale for model~C with this $\dot M$ are displayed in Figures~1b and
1c by thin-solid lines. This small reduction of $\dot M$ increases the
entropy per baryon only about a few percents. The dynamical timescale
also increases by a factor of $\lesssim 2$ for $L_\nu \le
10^{52}$~ergs~s$^{-1}$. For $L_\nu > 10^{52}$~ergs~s$^{-1}$, however,
the dynamical timescale increases drastically. This is because the winds
pass $r_{\rm c}$ with $T > 0.2$~MeV for higher luminosity
cases. Furthermore, the velocity and the radial gradient in temperature
significantly decrease after the wind crosses $r_{\rm c}$ for a subsonic
wind (Paper~I). This situation that the dynamical timescale becomes
significantly longer at an early stage is similar to the results of
previous detailed hydrodynamic simulations (\cite{Woos94};
\cite{Witt94}). Obviously, the $r$-process would not take place in the
winds for $L_\nu > 10^{52}$~ergs~s$^{-1}$ owing to the too long
dynamical timescales, as to be confirmed by nucleosynthesis calculations
in the next section. Note that the entropy per baryon for $L_\nu >
10^{52.6}$~ergs~s$^{-1}$ and the dynamical timescale for $L_\nu >
10^{52.2}$~ergs~s$^{-1}$ are not displayed in Figures~1a and 1b, since
the temperature at $10^4$~km, where the calculation is stopped, is
higher than 0.5~MeV and 0.2~MeV, respectively.

\section{THE $r$-PROCESS}

\begin{figure*}[t]
\epsscale{1.5}
\plotone{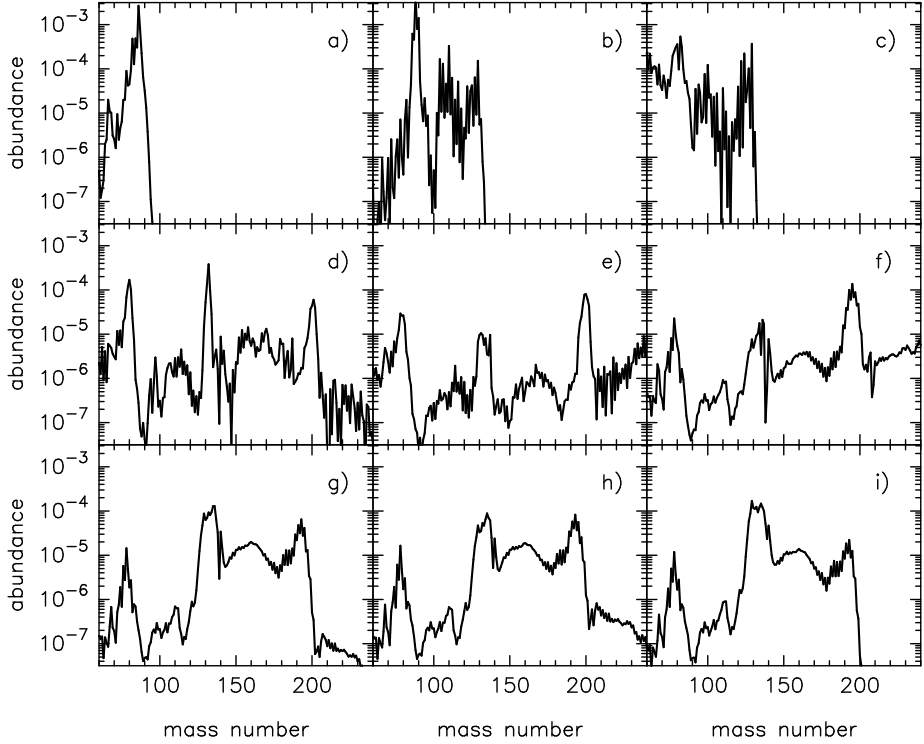}
\caption{\footnotesize Final abundances in model~C as a function of
atomic mass number, for $\log L_\nu$~(ergs~s$^{-1}) = $ 52.6 (a), 52.4
(b), 52.2 (c), 52.0 (d), 51.8 (e), 51.6 (f), 51.4 (g), 51.2 (h), 51.0
(i).}
\end{figure*}

Nucleosynthesis yields from the $r$-process were calculated for the
three neutrino-driven wind models discussed in the previous section.  We
used a fully implicit nuclear reaction network for both the
$\alpha$-rich freezeout and the subsequent $r$-process. The network
consists of over 3000 isotopes all the way from neutron and proton up to
the plutonium isotopes. We include all relevant reactions, i.e., $(n,
\gamma)$, $(p, \gamma)$, $(\alpha, \gamma)$, $(p, n)$, $(\alpha, p)$,
$(\alpha, n)$, and their inverse.  Reaction rates were taken from
Thielemann (1995) for nuclei with $Z \le 46$ and from Cowan, Thielemann,
\& Truran (1991) for those with $Z \ge 47$. Weak interactions such as
$\beta$-decay, $\beta$-delayed neutron emission, electron capture, and
the capture of electron and anti-electron neutrinos on free nucleons
were also included. These neutrino capture processes are of importance
as they reduce the number of free neutrons by the ``$\alpha$-effect''
(Meyer, McLaughlin, \& Fuller 1998). Note that fission reactions are not
included.

Nucleosynthesis calculations were carried out in nine trajectories with
neutrino luminosities between $\log L_\nu ({\rm ergs~s}^{-1}) = 52.6$
and 51.0 in intervals of 0.2~dex. The mass ejection rates were taken to
be $0.995 \times \dot M_{\rm max}$ as described in the previous section. 
Each calculation started when the temperature decreased to $T_9 = 9$
(where $T_9 \equiv T/10^9$~K). At this point the nuclear statistical
equilibrium (NSE) consists mostly of free nucleons. The initial mass
fractions of neutrons and protons were therefore given by NSE plus
charge neutrality: $X_n = 1 - Y_e$ and $X_p = Y_e$, respectively. The
initial electron fraction $Y_e$ was set to be 0.40, which is near the
lower bound for neutrino-heated ejecta from the models of Woosley et al. 
(1994). This approximation is adequate for our study, although
ultimately the initial electron fraction should be determined by the
detailed balance of the energy and luminosity between electron and
anti-electron neutrinos with a hydrodynamic study.

Figure~2 illustrates the final abundance of each trajectory for model~C
as a function of mass number. It is found that the trajectories with
$\log L_\nu ({\rm ergs~s}^{-1}) \le 52.0$ (Figure~2d-i) result in a
robust $r$-process. This is due to the sufficiently high entropies, $ S
\gtrsim 140 k$, and short dynamical timescales, $\tau \lesssim 30$~ms
(Figure~1). Figure~3 shows the neutron-to-seed ratio $Y_n/Y_{\rm seed}$
(solid line), the mass fraction of seed nuclei $X_{\rm seed}$ (dashed
line), and the electron fraction $Y_e$ (dot-dashed line) at $T_9 = 2.5$,
which is approximately the temperature at the beginning of the
$r$-process phase. Here, ``seed'' refers to all nuclei with $A \ge
12$. The final mass fraction of $r$-process elements ($A \ge 100$) $X_r$
is also shown by the dotted line. As can be seen in Figure~3, the
neutron-to-seed ratio reaches the maximum $\sim 160$ at $\log L_\nu
({\rm ergs~s}^{-1}) = 51.8$. As a result, for the corresponding
trajectory, the third abundance peak and the heavier are prominent
(Figure~2e). However, the total abundance of $r$-process nuclei produced
from this trajectory is small. This is because the mass fraction of seed
nuclei obtains the minimum value at this luminosity, as can be seen in
Figure~3. Note that $X_r$ is smaller for $L_\nu > 10^{52}$~ergs~s$^{-1}$
despite higher $X_{\rm seed}$ owing to the smaller neutron-to-seed
ratio. For lower luminosities $\log L_\nu ({\rm ergs~s}^{-1}) \le 51.6$,
$\beta$-delayed neutron emission smoothes the abundance pattern as seen
in Figure~2, because of the longer dynamical timescales (Figure~1).

\begin{figure}[t]
\epsscale{1.0}
\plotone{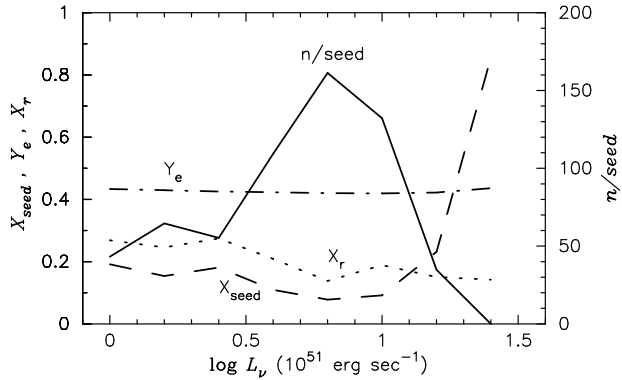}
\caption{\footnotesize The neutron-to-seed ratio (solid line), the mass
fraction of seed nuclei (dashed line), and the electron fraction
(dot-dashed line) at $T_9 = 2.5$. The final mass fraction of $r$-process
elements ($A \ge 100$) is also denoted by the dotted line.}
\end{figure}

As anticipated in the previous section, trajectories for $\log L_\nu \ge
52.2$ (Figures~2a-c) do not contribute significantly to the $r$-process,
because of the too long dynamical timescales (Figure~1c). Neutrino
capture on free nucleons somewhat increases the electron fraction to
$Y_e \sim 0.42-0.43$ at $T_9 = 2.5$ by the $\alpha$-effect
(\cite{Meye98}) as seen in Figure~3 (dot-dashed line). Thus, it is
important to include the neutrino capture reactions in order to
correctly calculate the resulting $r$-process nucleosynthesis. This
effect is not, however, as serious as that suggested by Meyer et al.
(1998) in the present $r$-process conditions. The reason is that the
dynamical timescales in our models $\tau \lesssim 30$~ms are
sufficiently shorter than the neutrino interaction lifetime for the
neutrino luminosities considered.  The dynamical timescale of Meyer et
al. (1998) was fixed at a value as large as $\tau = 0.3$~s.

In order to compare the nucleosynthesis results of each model with the
solar-system $r$-process abundances, the yields were integrated over the
mass-weighted time history in the following manner. The time evolution
of the neutrino luminosity is approximately given by
\begin{equation}\label{eqn:lnu}
L_\nu = L_{\nu 0} \exp (-t/\tau_\nu),
\end{equation}
where $L_{\nu 0}$ is the initial neutrino luminosity at the neutrino
sphere, and $\tau_\nu$ is the cooling timescale. We take $L_{\nu 0}$ to
be $\approx 4 \times 10^{52}$~ergs~s$^{-1}$ ($\log L_{\nu 0} = 52.6$)
from Woosley et al. (1994). In principle $L_{\nu 0}$ depends on the
radius of neutrino sphere and should be corrected for the gravitational
redshift effect on neutrino energy. However, the present approximation
is adequate for our purposes as neutrino-driven winds with $L_\nu
\gtrsim 4 \times 10^{52}$~ergs~s$^{-1}$ do not substantially contribute
to the $r$-process yields. The cooling timescale $\tau_\nu$ is set equal
to $1.25$~s so that the total time integrated neutrino energy becomes $3
\times 10^{53}$~ergs. With this $\tau_\nu$, the Figures~2a-i correspond
to time slices at 0, 0.6, 1.2, 1.7, 2.3, 2.9, 3.5, 4.0, and 4.6~s,
respectively.

\begin{figure}[t]
\epsscale{0.9}
\plotone{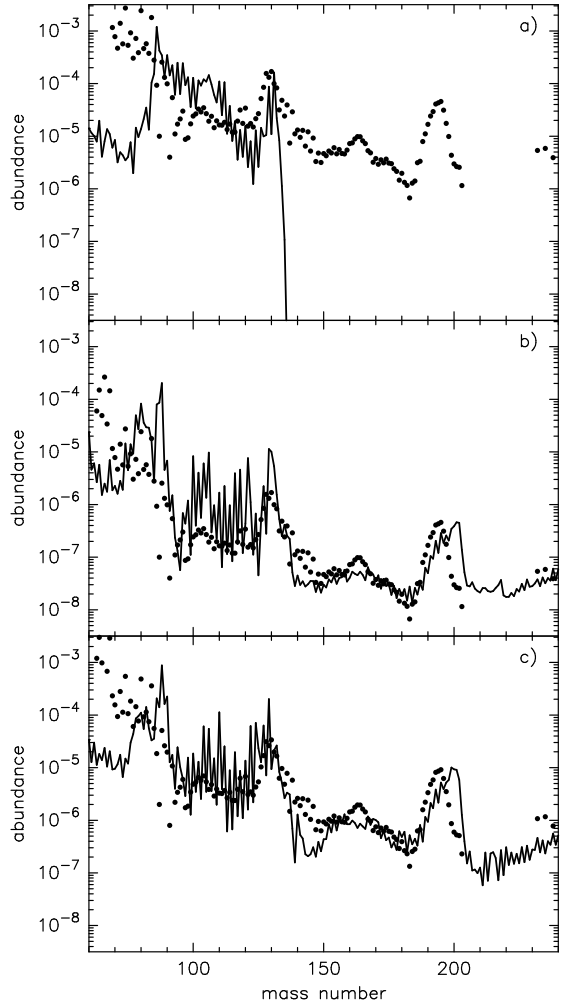}
\caption{\footnotesize The mass-weighted integrated yields for models~A
(a), B (b), and (c) as functions of mass number (lines). Also denoted
are the scaled solar $r$-abundances (points).}
\end{figure}

We integrate the mass-weighted $r$-process yields assuming that the
trajectories can be described at each time by steady flow corresponding
to the neutrino luminosity and neutron-star radius at that time,
\begin{equation}
Y_i = \frac{1}{M_{\rm ej}} \int Y_i (t) \dot M (t) dt,
\end{equation}
where
\begin{equation}
M_{\rm ej} = \int \dot M (t) dt
\end{equation}
is the total mass ejected from the neutron star.

Obviously, our assumption of steady flow for each trajectory is
oversimplified. However, the thermal properties of material in the
neutrino-heated wind are mostly determined near the surface of the
neutron star $\lesssim 30$~km, where the equilibration time ($\lesssim
0.1$~s) is shorter than the neutrino luminosity timescale $\tau_\nu$
($\sim 1$~s). In addition, during the relaxation of the neutron star
radius $\sim 2$~s, the neutrino luminosity is so high $L_\nu \gtrsim
10^{52}$~ergs~s$^{-1}$ that few $r$-process elements are
synthesized. Hence, the assumptions of steady flow for each neutrino
luminosity and a fixed neutron-star radius are probably reasonable.

In Figure~4, the integrated yields for models~A (a), B (b), and C (c)
are compared with the solar-system $r$-process abundances (K\"appeler,
Beer, \& Wisshak 1989). In models B and C the solar $r$-process
abundances are normalized to the third peak.  For model A the second
peak is used for normalization. Comparison of the three results suggests
that the ``compaction'' of the proto-neutron star is essential to
reproduce the pattern of the solar $r$-process abundances. Indeed, the
abundance pattern for elements with $A \gtrsim 120$ in models~B and C
are in excellent agreement with the scaled solar $r$-abundances, while
only a weak $r$-process occurred in model~A.

In all three models there is about a factor of 10 overproduction of
nuclei with $A \lesssim 120$. For $A \approx 90$ in particular, there is
a prominent peak as also seen in previous studies (\cite{Frei99a}). In
the present models, however, the overproduction is about a factor of 10
smaller than that of Woosley et al. (1994). Moreover, if the abundances
of $A \sim 100-130$ were smoothed, they would agree reasonably well with
the solar $r$-pattern.

\begin{figure}[t]
\epsscale{0.9}
\plotone{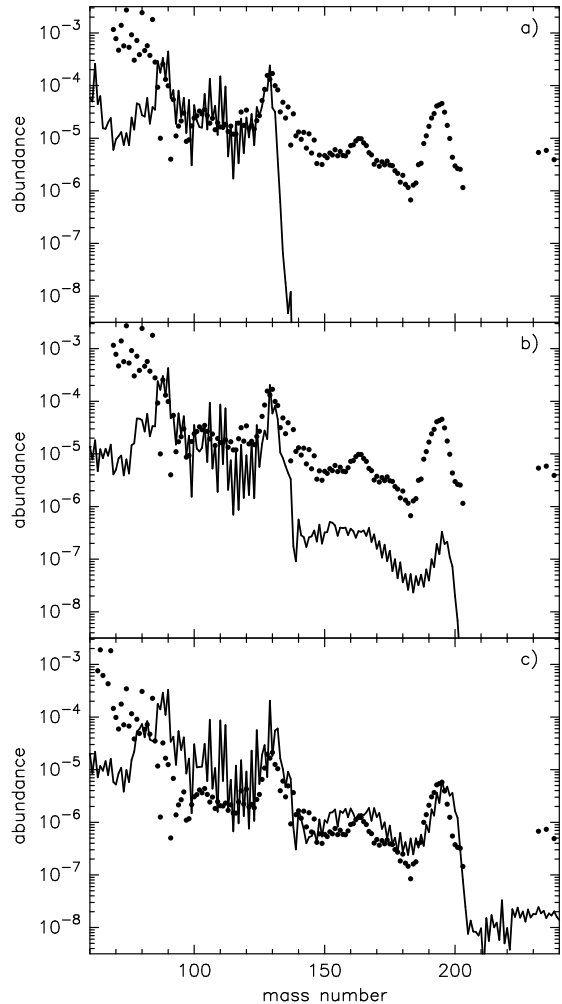}
\caption{\footnotesize Same as the Figure~4, but for $(M, R)$ $= (1.7
M_\odot, 10$~ km), $(1.8 M_\odot, 10$~km), and $(1.9 M_\odot, 10$~km).}
\end{figure}

The good agreement in models B and C is a consequence of the robust
$r$-process which starts at the early phase ($\sim 2$~s) of the
neutrino-driven wind. At this time $L_\nu$ is still as high as $\sim
10^{52}$~ergs~s$^{-1}$ and the integrated yields are dominated by the
matter ejected there. This is in contrast to the result of Woosley et
al. (1994) in which the $r$-process occurs only during the late phase of
the neutrino-driven wind ($\sim 10$~s), where the mass ejection rate has
already significantly declined.

\begin{deluxetable}{rrrrr}
\tablecaption{Total Ejected Mass \label{tab2}}
\tablewidth{0pt}
\tablehead{
\colhead{} & \colhead{A} & \colhead{B} & \colhead{C} &
\colhead{W94\tablenotemark{\dagger}}}
\startdata
$M_{\rm ej}/M_\odot$   & $2.9\times 10^{-3}$ & $1.7\times 10^{-3}$ &
$1.1\times 10^{-3}$ & $3.2\times 10^{-2}$ \\
$M_{{\rm ej,}\,r}/M_\odot$ & $5.3\times 10^{-4}$ & $2.0\times 10^{-5}$ &
$1.4\times 10^{-4}$ & $5.8\times 10^{-5}\tablenotemark{\ddagger}$ \\
\enddata
\footnotesize
\tablenotetext{\dagger}{Woosley et al. (1994)}
\tablenotetext{\ddagger}{abundances produced in the last 16 trajectories}
\end{deluxetable}

In Table~2, the total ejected mass $M_{\rm ej}$ and mass of $r$-process
($A \ge 100$) elements $M_{{\rm ej,}\,r}$ for each model are compared to
those of Woosley et al. (1994). Note that $M_{{\rm ej,}\,r}$ in the W94
column is the sum of the last 16 trajectories in Woosley et al. (1994),
whose pattern was in excellent agreement with the solar
$r$-abundances. The ejected $r$-process mass in model~C is about a
factor of two larger than that of Woosley et al. (1994) owing to
production of $r$-nuclei in early times when $\dot M$ is high. It is
interesting to note that the total ejected masses of all models are
smaller than that of Woosley et al. (1994), while $\dot M$ is set to be
nearly equal to the maximum value in this study. This is a consequence
of our more compact neutron stars, which obtain smaller $\dot M$ as can
be seen in Figure~1a. In particular, the radius of the neutron star in
Woosley et al. (1994) is much larger than 10~km in the first a few
seconds, while those in this study are taken to be constant. In
addition, $\dot M$ in Woosley et al. (1994) during the first $\sim 1$~s
is significantly larger than expected for a steady state wind.

Comparing the three models in Table~2 we see that $M_{\rm ej}$ does not
significantly change from model to model, being somewhat smaller in
models with a deeper gravitational potential. As can be seen in
Figure~4, however, the $r$-abundance pattern is substantially different
among these models. Model~A reproduces the first ($A \sim 80$) and
second ($A \sim 130$) $r$-process peaks, but no elements with $A\gtrsim
130$ are synthesized. In contrast, the solar $r$-pattern, including all
three $r$-process peaks, is well reproduced in the ``compact'' neutron
star models B and C. This is a consequence of the significantly higher
entropies $S \gtrsim 130 N_A k$ and shorter dynamical timescales $\tau
\lesssim 30$~ms for these models.

Comparing the absolute yields in models~B and C, we find that the
neutrino-driven wind of a more massive proto-neutron star produces more
$r$-process elements when the compaction ratio is the same. Indeed, the
mass fraction of $r$-elements in model~C is about a factor of 10 larger
than that of model~B. The reason is that the dynamical timescale in
model~B is shorter than in model~C (cf.~Figure~1). In such fast winds,
fewer seed nuclei are produced, although the neutron-to-seed ratio is
high enough for a robust $r$-process (see also Hoffman, Woosley, \& Qian
1997).

Let us discuss how sensitive these results are to the compaction
ratio. We present in Figure~5 the mass-weighted integrated yields for
three more cases $(M, R)$ $= (1.7 M_\odot, 10$~ km), $(1.8 M_\odot,
10$~km), and $(1.9 M_\odot, 10$~km). These correspond to compaction
ratios of 0.25, 0.27, and 0.28, respectively. Interestingly, the pattern
in the first model is not significantly different from that of
model~A. In fact, it is in slightly better agreement with the first and
second solar $r$-process peaks but does not produce heavier elements
(Figure~5a). The second model produces a small amount of $A > 130$
elements, and the third one reasonably reproduces the solar $r$-pattern
up to the third peak (Figures~5b and c). These calculations indicate
that the $r$-process strongly depends on the compaction ratio. A good
fit is only obtained when the compaction ratio is $\gtrsim 0.28$.

\section{DISCUSSIONS AND CONCLUSION}

\begin{figure}[t]
\epsscale{0.7}
\plotone{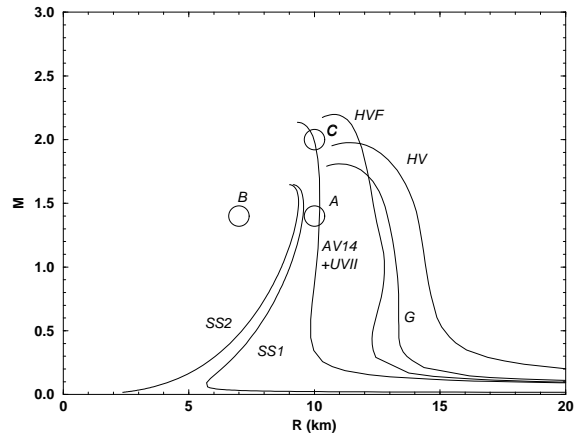}
\caption{\footnotesize Comparison of the mass radius relation (lines) for
various equations of state with the models (circles) A, B, and C
considered in this work. See text for an explanation of the various
equations of state.}
\end{figure}

We have explored the $r$-process in neutrino-heated ejecta
(neutrino-driven winds) from nascent neutron stars for various
neutron-star masses, radii, and neutrino luminosities. Our results
suggest that proto-neutron stars with a large compaction ratio ($GM/c^2R
\sim 0.3$; models~B and C), where general relativistic effects are
important, provide the best conditions for the $r$-process.  Such
compact proto-neutron stars lead to sufficiently high entropy $S \sim
100-200 N_A k$ and short dynamical timescales $\tau \lesssim 30$~ms to
produce an optimum neutron to seed ratio for material in the
neutrino-driven winds. The total time-integrated yields from these
models are in good agreement with the solar $r$-process abundance
pattern especially for nuclei with $A \sim 120-200$. Our short dynamical
timescale and assumed condition of steady-flow equilibrium for material
in the winds lead to a good $r$-process even at early times when the
neutrino luminosity is still high $L_\nu \sim 10^{52}$~ergs~s$^{-1}$. As
a result, the overproduction of nuclei $A \lesssim 120$ is significantly
reduced, compared to that of Woosley et al. (1994), although being still
overproduced by a factor of $\sim 10$.

Our results also suggest that, for models of the same compaction ratio,
e.g., B and C, a higher neutron-star mass yields more $r$-process
material. This is due to the somewhat longer dynamical timescale for
material ejected from high-mass neutron stars which increases the amount
of seed nuclei.

The implication of these two results is that an optimum $r$-process
which does not overproduce $A \sim 90$ nuclei may require both a well
developed bubble and a massive proto-neutron star while the $r$-process
elements are synthesized. It should be noted that we presume that the
bubble is already evacuated and a steady-state wind developed for all
neutrino luminosities. For the dynamical supernova model of Woosley et
al. (1994), however, the proto-neutron star is still relaxing and the
bubble slowly expanding during the first $\sim 2$~s. In the model~C of
this study, the $r$-process starts at $L_\nu = 10^{52}$~ergs (Figure~2),
which corresponds to 1.7~s in equation~(5). At $L_\nu \sim
10^{52}$~ergs, therefore, the entropy per baryon might be somewhat
overestimated, although the dynamical timescale as a function of $L_\nu$
shows a similar trend to those of previous hydrodynamic studies
(Figure~1c). A detailed hydrodynamic study of the neutrino-induced
explosion for a massive supernova progenitor will be eventually needed
to confirm whether the early stage of the neutrino-driven winds from
compact neutron stars are really the astrophysical site of the
$r$-process.

Alternatively, this scenario might occur if the low entropy material
initially ejected from the supernova were to fall back onto the neutron
star. This would produce the massive remnant. This epoch might then
follow by a second epoch of high neutrino flux into the previously
evacuated bubble. This might occur, for example, if the proto-neutron
star were to experience a late transition to strange matter and
subsequent neutrino emission or a late softening and heating by kaon
condensation (Thorsson, Prakash, \& Lattimer 1994). For example, the
softening of a proto-neutron star by kaon condensation is estimated to
occur in a few to 10 seconds after core bounce (\cite{Keil95}), which
may be relevant to the $r$-process in this study. In either case, it is
clear that these new results could suggest an interesting new twist to
the supernova paradigm.

If these models for the $r$-process are correct, then our results could
also pose a significant constraint on the equation of state for the
neutron-star remnant. A summary of the mass-radius relationship of some
contemporary equations of state is shown in Figure~6. These are compared
with the various models considered in this work. These representative
equations of state include an example of a non-relativistic nucleon
potential model equation of state with both two-body and three-body
terms (AV14 + UVII; Wiringa, Fiks, \& Fabrocini 1988) as well as various
relativistic mean field equations of state, including those based upon a
relativistic Hartree (G; Glendenning 1989 and HV; Glendenning 1985),
Hartree Fock (HVF; Weber \& Weigel 1989) calculation and those with
strange-matter interiors (SS1 \& SS2; Glendenning \& Weber 1992).

Model~A with $M = 1.4 M_\odot$ and $R = 10$~km is a typical neutron star
mass and a typical radius for a non-relativistic EOS and several
relativistic ones as well. However, it does not produce a good
$r$-process in our simulations.  Generating the observed abundance curve
seems to require a slightly more exotic paradigm.  If we consider our
best $r$-process model, i.e. model~C with $M = 2.0 M_\odot$ and $R =
10$~km, we see that many of the equations of state lead to such a
remnant.  In all cases, however, such a massive remnant is very close to
or equal to the maximum mass allowed by the EOS.  If such a remnant is
required to produce the $r$-process, then it seems that the $r$-process
must occur for a remnant that has accreted to its maximum mass and is
about to collapse to a black hole. This suggests another possible
paradigm in which the $r$-process occurs in the previously evacuated
bubble by neutrinos generated as the short-lived proto-neutron star
begins its collapse to a black hole (see also Qian, Vogel, \& Wasserburg
1998). This would perhaps require that the $r$-process occurs in a more
massive progenitor star which does not leave a neutron star remnant such
as probably occurred in SN 1987A (\cite{Brow94}).

Although model~B with $M = 1.4 M_\odot$ and $R = 7$~km also gives a
reasonable $r$-process abundance distribution, it is very difficult to
form such a compact neutron star of this mass.  Such lower-mass high
compaction stars could perhaps form if a strange matter core develops as
in SS1 and SS2 of Figure~6.

The results of this study are also of importance for the chemical
evolution of the Galaxy. Ishimaru \& Wanajo (1999) have suggested that
the large dispersion of europium with respect to iron among metal-poor
stars in the Galactic halo (\cite{Ryan96}; \cite{McWi97}) can be
reproduced if the $r$-process elements originate from stars of $\ge 30
M_\odot$ or $8-10 M_\odot$. Our results also imply that, for a given
equation of state, a more massive proto-neutron star leads to a smaller
compaction ratio and therefore more favorable conditions for the
$r$-process.  Furthermore, if the maximum mass of a proto-neutron star
were as large as $\sim 2 M_\odot$, the progenitor star is probably very
massive. Such stars with $\ge 30 M_\odot$ account for only $\sim 10 \%$
of all supernova progenitors with a typical initial mass
function. Nevertheless, the amount of $r$-process matter in model~C is
about a factor of 2 higher than that of Woosley et al. (1994), so that
such a restriction to high mass stars may be reasonable.

It should be noted that, even if neutrino-driven winds fail to reproduce
the solar $r$-pattern especially of the third peak, they still could be
significant sources of the solar $r$-elements of at least the second
peak and lighter. If that is the case, heavier $r$-process elements must
be synthesized in other sources. This is consistent with the scenario of
multiple $r$-process sites implied by meteoritic abundances (Wasserburg,
Busso, \& Gallino 1996), spectroscopic studies of metal-poor halo stars
(\cite{Sned00}), and the chemical evolution studies
(\cite{Ishi00}). This could be also explained by the neutrino-driven
wind models described here, which show the strong dependence of the
$r$-process yields on the remnant mass.

In fact, most of the metal-poor stars in the Galactic halo show the
higher ratios of light-to-heavy $r$-elements (e.g., Sr/Ba or Sr/Eu) than
that of the solar $r$-abundances, although the values differ from star
to star by up to $\sim 2$ orders of magnitude (\cite{McWi98};
\cite{Ishi00}). This implies that these $r$-elements might originate
from the ``weak'' $r$-processing associated with smaller compaction
ratios $< 0.28$ (Figures~4 and 5). However, some metal-poor stars (e.g.,
CS 22892-052) show somewhat smaller ratios of light-to-heavy
$r$-elements compared to that of the solar $r$-abundances, by a factor
of $\sim 2-3$ (Sneden et al. 2000; see also Wasserburg et al. 1996 for a
meteoritic study), which cannot be reproduced by any models in this
study. If such low light-to-heavy ratios originate from neutrino-driven
winds, we need somewhat more compact proto-neutron stars than considered
in this study, otherwise we have to seek other causes to reduce the
lighter $r$-nuclei, e.g., fallback of the first, low entropy material
ejected (\cite{Woos94}).

Obviously, our approach of assuming constant values for the initial
electron fraction, neutrino luminosity in each trajectory, the radius of
the proto-neutron star, as well as an exponential time evolution of
neutrino luminosity, is too simplified to conclude that these
neutrino-driven wind models are an accurate description of the
$r$-process.  For example, the time variation of $Y_e$ is of particular
importance in the $r$-process. Recent hydrodynamic studies of
core-collapse supernovae with accurate treatments of the neutrino
transport have shown that $Y_e$ exceeds 0.5 for the first a few hundred
ms after core bounce (\cite{Ramp00}). If that is the case, then the
overproduction of $A \approx 90$ nuclei in our results would disappear
(\cite{Hoff96}). In the present study, the initial electron fraction was
fixed at a relatively low value, 0.4, although it increases to
$0.42-0.43$ by the time of onset of the $r$-process. However, $Y_e$ may
be significantly higher than that during the early phase of the
neutrino-driven wind. Even with a higher $Y_e$, however, a sufficient
neutron-to-seed ratio could be obtained if the $\alpha$-process is less
efficient (\cite{Hoff97}). Instead, the yields of $r$-process elements
might decrease owing to a smaller amount of seed material.

Future hydrodynamic studies of core-collapse supernovae with accurate
neutrino transport, including multidimensional effects such as
convection and rotation, as well as magnetic fields will probably
ultimately be required before we can be certain that the neutrino-driven
winds are indeed the major $r$-process site in the Galaxy. Nevertheless,
our results have confirmed that the scenario of neutrino-driven winds is
still viable and promising as the true astrophysical site for the
$r$-process.

\acknowledgments

We would like to acknowledge useful discussions with Y. Ishimaru,
N. Itoh, K. Sumiyoshi, and M. Terasawa. This work was supported in part
by Japan Society for Promotion of Science, and by the Grant-in Aid for
Scientific Research (1064236, 10044103, 11127220, and 12047233) of the
Ministry of Education, Science, Sports and Culture of Japan, and DoE
Nuclear Theory grant DE-FG02-95-ER40394.


\begin{thebibliography}{}

\small

\bibitem[Baumgarte et al. 1996]{Baum96}
 Baumgarte, T. W., Janka, H. -T., Keil, W., Shapiro, S. L.,
 \& Teukolsky, S. A. 1996, \apj, 468, 823
\bibitem[Brown \& Bethe 1994]{Brow94}
 Brown, G. W. \& Bethe, H. A. 1994, \apj, 423, 659
\bibitem[Cardall \& Fuller 1997]{Card97}
 Cardall, C. Y. \& Fuller, G. M. 1997, \apjl, 486, L111
\bibitem[Cowan, Thielemann, \& Truran 1991]{Cowa91}
 Cowan, J., Thielemann, F. -K., \& Truran, J. W. 1991, \physrep, 208, 267
\bibitem[Freiburghaus et al. 1999]{Frei99a}
 Freiburghaus, C., Rembges, J. -F., Rauscher, T., Kolbe, E.,
 Thielemann, F. -K., Kratz, K. -L., Pfeiffer, B., \& Cowan, J. J.
 1999, \apj, 516, 381
\bibitem[Freiburghaus et al. 1999]{Frei99b}
 Freiburghaus, C., Rosswog, S., \& Thielemann, F. -K. 1999, \apjl, 525, L121
\bibitem[Glendenning 1985]{Glen85}
 Glendenning, N. K. 1985, \apj, 293, 470
\bibitem[Glendenning 1989]{Glen89}
 Glendenning, N. K. 1989, \nphysa, 493, 521
\bibitem[Glendenning \& Weber 1992]{Glen92}
 Glendenning, N. K. \& Weber, F. 1992, \apj, 400, 647
\bibitem[Hoffman et al. 1996]{Hoff96}
 Hoffman, R. D., Woosley, S. E., Fuller, G. M., \& Meyer, B. S. 1996,
 \apj, 460, 478
\bibitem[Hoffman et al. 1997]{Hoff97}
 Hoffman, R. D., Woosley, S. E., \& Qian, Y. -Z. 1997, \apj, 482, 951
\bibitem[Ishimaru \& Wanajo 1999]{Ishi99}
 Ishimaru, Y. \& Wanajo, S. 1999, \apjl, 511, L33
\bibitem[Ishimaru \& Wanajo 2000]{Ishi00}
 Ishimaru, Y. \& Wanajo, S. 2000, in {\it First Stars}, A. Weiss,
 T. Abel,\& V. Hill eds. Springer-Verlag, Berlin, p. 189
\bibitem[Itoh, Hayashi, Nishikawa, \& Kohyama 1996]{Itoh96}
 Itoh, N., Hayashi, H., Nishikawa, A., \& Kohyama, Y. 1996, \apjs, 102, 411
\bibitem[K\"appeler et al. 1989]{Kapp89}
K\"appeler, F., Beer, H., \& Wisshak, K. 1989, Rep. Prog. Phys., 52, 945
\bibitem[Keil \& Janka 1995]{Keil95}
Keil, W. \& Janka, H. -Th 1995, \aap, 296, 145
\bibitem[McWilliam 1997]{McWi97}
 McWilliam, A. 1997, \araa, 35, 503
\bibitem[McWilliam 1998]{McWi98}
 McWilliam, A. 1998, \aj, 115, 1640
\bibitem[Meyer et al. 1992]{Meye92}
 Meyer, B. S., Mathews, G. J., Howard, W. M., Woosley, S. E.,
 \& Hoffman, R. D. 1992, \apj, 399, 656
\bibitem[Meyer et al. 1998]{Meye98}
 Meyer, B. S., McLaughlin, G. C., \& Fuller G. M. 1998, \prc, 58, 3696
\bibitem[Otsuki et al. 2000]{Otsu00}
 Otsuki, K., Tagoshi, H., Kajino, T., \& Wanajo, S. 2000, \apj, 533, 424
 (Paper~I)
\bibitem[Qian \& Woosley 1996]{Qian96}
 Qian, Y. -Z. \& Woosley, S. E. 1996, \apj, 471, 331
\bibitem[Qian et al. 1998]{Qian98}
 Qian, Y. -Z., Vogel, P., \& Wasserburg, G. J. 1998, \apj, 494, 285
\bibitem[Qian 2000]{Qian00}
 Qian, Y. -Z. 2000, \apjl, 534, L67
\bibitem[Rampp \& Janka 2000]{Ramp00}
 Rampp, M. \& Janka, H. -T. 2000, \apjl, 539, L33
\bibitem[Ryan et al. 1996]{Ryan96}
 Ryan, S. G., Norris, J. E., \& Beers, T. C. 1996, \apj, 471, 254
\bibitem[Shapiro \& Teukolsky 1983]{Shap83}
 Shapiro, P., \& Teukolsky, S. 1983, Black Holes, White Dwarfs, and
 Neutron Stars (New York: Wiley)
\bibitem[Sneden et al. 1996]{Sned96}
 Sneden, C., McWilliam, A., Preston, G. W., Cowan, J. J., Burris, D. L.,
 \& Armosky, B. J. 1996, \apj, 467, 819
\bibitem[Sneden et al. 1998]{Sned98}
 Sneden, C., Cowan, J. J., Debra, L. B., \& Truran, J. W. 1998, \apj, 496,
235
\bibitem[Sneden et al. 2000]{Sned00}
 Sneden, C., Cowan, J. J., Ivans, I. I., Fuller, G. M., Burles, S.,
 Beers, T. C., Lawler, J. E. 2000, \apjl, 533, L139
\bibitem[Sumiyoshi et al. 2000]{Sumi00}
 Sumiyoshi, K., Suzuki, H., Otsuki, K., Terasawa, M.,
 \& Yamada, S. 2000, \pasj, 52, 601
\bibitem[Takahashi et al. 1994]{Taka94}
 Takahashi, K., Witti, J., \& Janka, H. -Th. 1994, \aap, 286, 857
\bibitem[Thielemann 1959]{Thie95}
 Thielemann, F.-K. 1995, private communication
\bibitem[Thorsson et al. 1994]{Thor94}
 Thorsson, V., Prakash, M., \& Lattimer, J. M. 1994, \nphysa, 572, 693
\bibitem[Tsujimoto et al. 2000]{Tsuj00}
 Tsujimoto, T., Shigeyama, T., \& Yoshii, Y. 2000, \apjl, 531, L33
\bibitem[Wasserburg et al. 1996]{Wass96}
 Wasserburg, G. J., Busso, M., \& Gallino, R. 1996, \apjl, 466, L109
\bibitem[Weber \& Weigel 1989]{Webe89}
 Weber, F. \& Weigel, M. K. 1989, \nphysa, 505, 779
\bibitem[Wiringa et al. 1988]{Wiri88}
 Wiringa, R. B., Fiks, U., \& Fabrocini, A. 1988, \prc, 38, 1010
\bibitem[Wheeler\, Cowan\, \& Hillebrandt 1998]{Whee98}
 Wheeler, J. C., Cowan, J. J., \& Hillebrandt, W. 1998, \apjl, 493, L101
\bibitem[Witti et al. 1994]{Witt94}
 Witti, J., Janka, H. -Th., \& Takahashi, K. 1994, \aap, 286, 841
\bibitem[Woosley \& Hoffman 1992]{Woos92}
 Woosley, S. E. \& Hoffman, R. D. 1992, \apj, 395, 202
\bibitem[Woosley et al. 1994]{Woos94}
 Woosley, S. E., Wilson, J. R., Mathews, G. J., Hoffman, R. D., \&
 Meyer, B. S. 1994, \apj, 433, 229

\end{thebibliography}
\end{document}